\documentclass[10pt,final,journal,twocolumn]{IEEEtran}
\usepackage[T1]{fontenc}

\ifCLASSINFOpdf
\else
\fi
\usepackage{cite}
\usepackage{graphicx}
\usepackage{subfigure}
\usepackage{caption}
\usepackage[cmex10]{amsmath}
\usepackage{amssymb}
\usepackage{amsfonts}
\usepackage{tabularx}
\usepackage{multirow}
\usepackage{bigstrut}
\usepackage{booktabs}
\usepackage{makecell}
\usepackage{algorithm}
\usepackage{algorithmic}
\usepackage[caption=false,font=footnotesize]{subfig}
\usepackage{color}

\usepackage{mdwmath}
\usepackage{mdwtab}
\usepackage{eqparbox}
\usepackage{array}

\begin{document}
%

\title{Context-aware Group Buying in Ultra-dense Small Cell Networks: Unity is Strength}
%
%
%

\author{Yuli Zhang,~\IEEEmembership{Student Member,~IEEE,}
        Yuhua Xu,~\IEEEmembership{Member,~IEEE,}
        Alagan Anpalagan,~\IEEEmembership{Senior Member,~IEEE}
        \\
        Qihui Wu,~\IEEEmembership{Senior Member,~IEEE,}
        Yitao Xu,
        Youming Sun,
        Shuo Feng, Yunpeng Luo

\thanks{Yuli Zhang, Yuhua Xu, Yitao Xu, Youming Sun and Yunpeng Luo are with the College of Communications Engineering, Army Engineering University of the PLA, Nanjing, China, 210007. Yuli Zhang and Yuhua Xu are also with Science and Technology on Communication Networks Laboratory, Shijiazhuang 050002, China. Yuhua Xu is the corresponding author. (e-mail: yulipkueecs08@sina.com, xuyuhua365@163.com, yitaoxu@126.com, sunyouming10@163.com, 349492881@qq.com.)}
\thanks{Alagan Anpalagan is with the Department of Electrical and Computer Engineering, Ryerson University, Toronto, Canada (e-mail:alagan@ee.ryerson.ca).)}
\thanks{Qihui Wu is with College of Electronic and Information Engineering, Nanjing University of Aeronautics and Astronautics, Nanjing, 211106, China (email: wuqihui2014@sina.com.)}
\thanks{Shuo Feng is with Cognitive Systems Laboratory, McMaster University, Hamilton, ON L8S 4L8, Canada (e-mail: fengs13@mcmaster.ca)}
}

%
%

\markboth{Journal of \LaTeX\ Class Files,~Vol.~14, No.~8, August~2015}%
{Shell \MakeLowercase{\textit{et al.}}: Bare Demo of IEEEtran.cls for IEEE Communications Society Journals}
%



\maketitle

\begin{abstract}
The ultra-dense small cell networks (SCNs) have been regarded as a promising technology to solve the data traffic explosion in future. However, the complicated relationships among large scale users and cells practically demand a cooperative grouping mechanism to account for the shortage of resources in SCNs. Then, a group buying market mechanism provides a win-win situation and an effective resource allocation in the view of economics. With additional decision information being provided, the context awareness enhances and improves the group buying mechanism. In this article, we propose a Context-Aware Group Buying (CAGB) mechanism to allocate the resources in ultra-dense SCNs. The feasibility, necessity, and effectiveness of CAGB are analyzed first. Then, we introduce the group buying mechanism and some common context awareness. The relationship between context awareness and group buying is also analyzed. Some important technologies in SCNs are discussed in the view of CAGB, such as load balancing, spectrum management, and cooperative caching. Then, the graphical coalition formation games (CFGs) of CAGB are also presented to match the complicated network topologies in SCNs. Two CAGB-based use cases about spectrum market and cooperative caching are presented. Finally, future research issues about context awareness, grouping mechanism, and buying mechanism are discussed.
\end{abstract}


%
\IEEEpeerreviewmaketitle

\section{Introduction}

\IEEEPARstart{O}{ver} the past decade, the mobile communication devices and mobile data traffic have experienced an explosive growth, which exacerbates the spectrum shortage. In order to solve this problem and accommodate the heavy traffic, 5G technologies have been proposed to achieve a thousand times increase in throughput. As one of the most important features and promising technologies, the hype dense small cell networks (SCNs) \cite{Gexiaohu1} have drawn a lot of attention recently.

The ultra-dense property brings about the large number of cells and users firstly. Compared with the traditional communication systems, numerous players challenge the decision framework and learning algorithms in SCNs. The interactions among users and cells are more complicated and difficult to be modeled, such as in the spectrum market. Traditional distributed decision framework without information exchange and cooperation cannot handle the large scales situation well. But the connections among users and cells are stronger, especially for the closely deployed ones. These close connections of users' demands, contents, traffic or other aspects are considered as context awareness. Also, these connections require a cooperative mechanism to explore and use. Furthermore, the existing studies mostly focused on one level's resource allocation, lacking of an overall consideration for both of users and cells. Hence, a cooperative mechanism which uses multiple dimension information and achieves a win-win resource allocation is to be investigated.

In this article, a context-aware group buying (CAGB) mechanism is proposed to meet the resource allocation in ultra-dense small cell networks. Group buying is a traditional cooperative economic behavior and has been widely employed on the Internet, such as in eBay and Taobao. Reviewing some existing technologies in ultra-dense SCNs, such as spectrum resource allocation and management \cite{Gaolin,Groupon,SunJSAC,YangdejunTVT}, and load balancing \cite{Loadbalance,ZhangruiJSAC}, it can be seen that there are mainly three reasons that make group buying employed in ultra-dense small cell networks possible and effective:
\begin{itemize}
  \item The hyper-dense deployment provides a probability that we can seek the consistency of users' demands from the diversity. The products of traditional group buying are always hot sales, which attract people to form a group and buy them. This consistency or same demand of buyers is the basic motivation of group buying. In wireless communications, the spectrum, energy or other resources are hot sales for user, equipment and cells. To achieve these resources, players would like to unite and form a group.
  \item The group mechanism can achieve a win-win in economics. In traditional group buying, the common cost of products is reduced or shared by a large number of buyers. In this way, buyers get the products with lower price and gain benefit. On the other hand, even though the unit price is less, the final benefit of sellers still increases due to the increasing number of buyers. Also, in wireless communications, some common or basic costs can be reduced and a win-win result motivates both sellers and buyers to employ group buying mechanism.
  \item The group buying mechanism can improve the resource efficiency through uniting users together. In the grouping mechanisms, distributed powers can be united as an entirety and achieve something they could not when separated. As a cooperation mechanism, users help each other. And the group buying may inspire more potential users through reducing common cost sometimes. Moreover, considering context awareness which provides extra information and strategy dimension, the context-aware group buying mechanism could be better.
\end{itemize}
Therefore, combining context-aware group buying and ultra-dense networks together is feasible, and it would achieve better performance in resource allocation.

The remainder of this paper is organized as follows. The origin and application of the group buying mechanism is introduced and the context awareness of the SCNs is discussed in Section II. The existing studies of CAGB for resource allocation are reviewed and some technologies are investigated in Section III. With these specific technologies, graphical coalition formation games are presented in Section IV. In Section V, two use cases of spectrum auction and cooperative caching with the CAGB mechanism are presented. Finally, future research issues and the conclusion are given.

\section{Context-aware Group Buying}
\subsection{Group Buying}
Group buying is a traditional mechanism for sales promotion, which shows a new life with Internet development. This mechanism has become an important and necessary part of the Internet economics with effective results. Almost all popular e-shops have employed this mechanism to improve their sales. The main advantage of group buying is to save the public cost with a large number of participants. On one hand, the cost is reduced and buyers can get a lower price. On the other hand, although the price is less than the non-grouping situation, the increasing number of buyers leads to a total increase in the final benefit. In this way, group buying brings about a win-win situation for both buyers and sellers.

\begin{figure*}[!t]
\centering
\includegraphics[width=14cm]{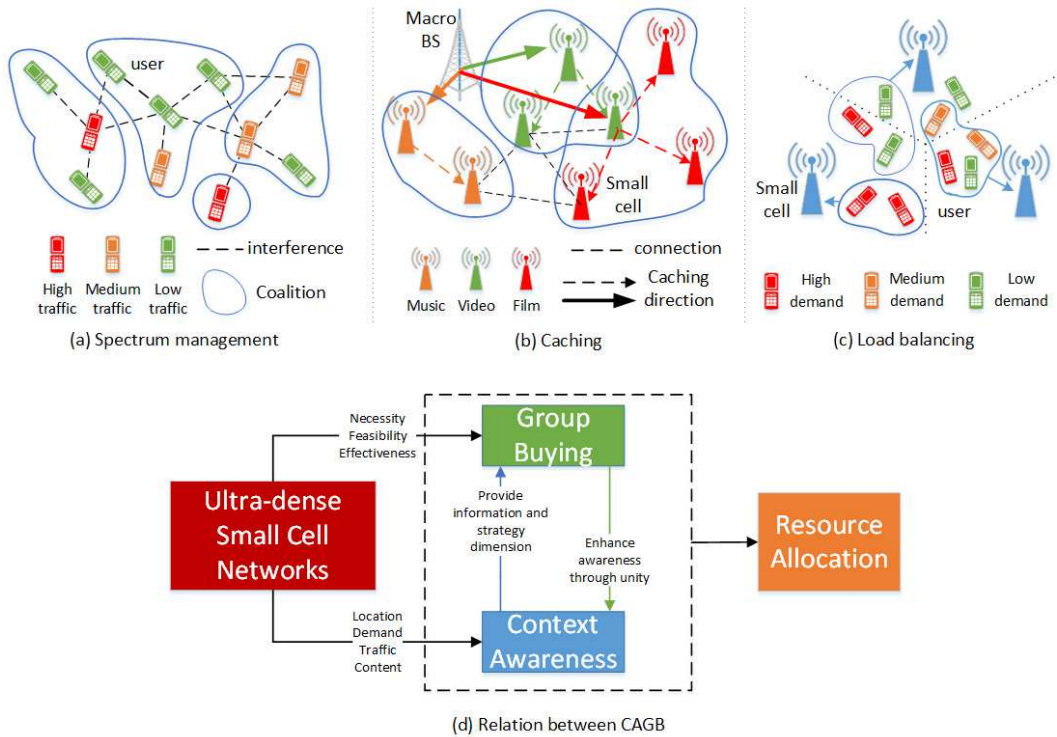} 
\caption{The context-aware group buying mechanism in SCNs and the relationship within CAGB.}
\label{threeefficiency}
\end{figure*}

There are already some researchers focusing on the group buying mechanism. Some typical models of spectrum auction, caching and load balancing about CAGB are shown in Fig. \ref{threeefficiency}. The group buying was first used in spectrum auction situations \cite{Groupon,YangdejunTVT}. They \cite{Groupon} investigated an auction pricing problem to achieve high benefit with keeping truthful property. Authors in \cite{YangdejunTVT} focused on the low budget situations and proposed a grouping rule to unite the low budget users. The low budget users are united based on their traffic demand, in order to increase the total budget and the final channel access probability. Then, the group buying was extended to other fields of small cell networks \cite{ZhangruiJSAC,WCL2017}. Authors in \cite{ZhangruiJSAC} studied the energy efficiency of SCNs under the energy market and an ahead selling framework. A Stackerlberg game was modeled with a group buying mechanism to balance the load between cells' energy resource and traffic demand. Our previous work \cite{WCL2017} studied a caching cost reduction problem and designed a group buying mechanism to reduce the data caching cost based on the content awareness.

Compared with the Internet situation, the group buying in ultra-dense SCNs has some differences and also, the key points,  which should be considered if employed.
\begin{itemize}
  \item The resource allocation is different from product sales. The main purpose of CAGB in SCNs is to optimize resource allocation and improve efficiency, instead of achieving more sales. Also, resources cannot be modeled into traditional products due to the heterogeneous, dynamic, uncertain and open properties of wireless communications.
  \item More complicated grouping mechanisms need to be designed. When optimized objectives are not similar specific products, users' utilities, grouping rules and results performance may be different. Then interaction among players also needs to be considered and formulated, which is totally different from traditional situations. Sometimes, users will play the role of the leaders in CAGB.
  \item A win-win result needs to be carefully designed. For the simple traditional group buying, the final win-win result is easy to achieve. While for the resource allocation, the results cannot be guaranteed to be win-win due to the redesigned group rules and complicated situations. The benefit of grouping sometimes belongs to only the buyer side while the sellers are ignored.
\end{itemize}

Most of the existing related work focused most attention on the buying actions instead of grouping strategies. However, we think that the group buying mechanism should go beyond its naming ``buying'' and, extend to more generalized situations with the core idea ``grouping''. In our opinion, the context-aware group buying is not only an economical benefit improvement tool, but also an important and efficient mechanism for resource allocation. For ultra-dense SCNs, the resource allocation and optimization with a large number of players are the challenges. Cell deployment \cite{DeploymentAccess}, spectrum management\cite{Gaolin,Groupon}, energy efficiency \cite{ZhangruiJSAC,Gexiaohu2} and caching \cite{WCL2017,Caching2} are the active research fields. Besides spectrum management, other fields are not related to the buying directly. However, the cooperative optimization idea can still be used in all these fields, especially when the distributed optimization has shown its disadvantages in ultra-dense situations. To solve this problem, finding the corresponding context awareness, designing the group forming rules and analyzing the performance are the three important folds to improve resource efficiencies.

\subsection{Context Awareness}
Based on the above discussion, to combine the group buying mechanism and resource allocation harmoniously, a context awareness bridge which can provide additional information and extend cooperation dimensions and strategy space is needed. The context awareness means that users can sense the environment and obtain the information, such as location and traffic,  achieve the information and improve their utilities driven by these awareness. According to the contents and the purpose, the context awareness can be classified into many categories. Here are some typical awareness:
\begin{itemize}
  \item Location awareness. In ultra-dense SCNs, cells and users are located closely in the hot spot area. Under this condition, some technologies can be realized, such as, device-to-device communications and load balancing. The location awareness is the most common among context awareness and influences the interference, throughput, delay and many other metrics in wireless communications.
  \item Content awareness. The content awareness is related to the caching technology or content dissemination networks. For these situations, what users concern about or focus on is very important for the service provider. With the deep learning in big data, users' favor or disfavor can be learned from their traffic data. Then, a more targeted service can be provided with better quality of service in more effective ways.
  \item Demand awareness. The QoS metric is the ability that system can provide, offer or guarantee for users. While, the quality of experience (QoE) is a complicated metric to evaluate service, including received service, user's environment influence, emotional state, and other issues. In this article, users' demand can be regarded as the simplification of the QoE metrics, which is the traditional QoS-level based on users' specific requirement. It can be bandwidth, throughput and other resource demand based on the application-level experiences. Compared with the traditional QoS metric, the demand reflects users' need more accurately. Driven by demand awareness, users' quality of experience will be improved. The resources can be allocated more effectively. On the other hand, the resource is saved by accurate demand, and therefore the saved part can serve more users.
  \item Energy awareness. Since the ultra-dense SCNs are developed recently, the large number of cells and users makes energy efficiency prominent due to the environmental and economic considerations. Cell sleep mode and energy harvesting technologies are proposed to save energy and achieve energy efficiency.
  \item Social awareness. Social aware networks have drawn a lot of attention recently. Besides physical interference and connection, users have more complicated social relationship. For example, two users may be friends and they would help each other through providing relay, extra spectrum resource and so on. Social awareness provides a social presentation of network topology.
\end{itemize}

Note that for almost all context-aware group buying, no matter with direct cooperation or indirect distributed coordination, a grouping mechanism is the key to design utilities and improve performance in both the user level and the network level. Imagining the following situation, if only one user is driven by its context awareness, such as demand, and others are just focusing on their throughput, the final result will not be better. Players need to unite and work together based on the context awareness. The relationship between context awareness and the group buying mechanism is given as follows, shown in Fig. \ref{threeefficiency}(d): (i) context awareness provides extra information dimension and strategy space, and (ii) group buying enhances the context awareness through unity and aggregation. The context awareness and group buying complement each other and the combination contributes to the resource allocation in SCNs.

\section{SCN Resource Allocation with CAGB}
In this section, we focus on some existing technologies in ultra-dense SCNs and try to find the connection with context-aware group buying, summarized as in Table I.

\begin{table*}[t]
\begin{center}
\textcolor[rgb]{0.00,0.07,1.00}{\caption{Resource allocation in SCNs}}
\begin{tabular}{|p{0.16\textwidth}|p{0.08\textwidth}|p{0.15\textwidth}|p{0.14\textwidth}|p{0.38\textwidth}|}
\hline
\makecell[c]{\textbf{Situations}} & \makecell[c]{\textbf{Literature}} & \makecell[c]{\textbf{Context Awareness}}  & \makecell[c]{\textbf{Strategy}} &  \makecell[c]{\textbf{Models}}  \\
\hline
\makecell[c]{\multirow{3}[6]{*}{Spectrum Management}} & \makecell[c]{\cite{SunJSAC}}  & \makecell[c]{Location + Traffic} & \makecell[c]{Channel} & \makecell[c]{Spectrum auction + overlapping coalition formation}\\
\cline{2-5}
& \makecell[c]{\cite{YangdejunTVT}} & \makecell[c]{Demand } & \makecell[c]{Coalition selection} &  \makecell[c]{Spectrum auction + coalition formation} \\
\cline{2-5}
& \makecell[c]{\cite{Gaolin}} & \makecell[c]{Location + Demand} & \makecell[c]{Payoff }& \makecell[c]{Two-side market}  \\
\cline{2-5}
& \makecell[c]{\cite{Sunlingyang}} & \makecell[c]{Location}  & \makecell[c]{Channel }& \makecell[c]{Interference management + Overlapping coalition games} \\
 \hline
\makecell[c]{\multirow{2}[6]{*}{Load Balancing}} & \makecell[c]{\cite{ZhangruiJSAC}}  & \makecell[c]{Traffic + Energy} & \makecell[c]{Energy } & \makecell[c]{Energy efficiency + Stackelberg game}\\
\cline{2-5}
& \makecell[c]{\cite{Loadbalance}} & \makecell[c]{Content } & \makecell[c]{User association} &  \makecell[c]{Spectrum auction + coalition formation} \\
\cline{2-5}
& \makecell[c]{\cite{Sleepmode}} & \makecell[c]{Energy } & \makecell[c]{Sleep mode }& \makecell[c]{Local cooperative game} \\
 \hline
\makecell[c]{\multirow{2}{*}{Caching}} & \makecell[c]{\cite{WCL2017}}  & \makecell[c]{Content } & \makecell[c]{Caching source } & \makecell[c]{Local cooperative game}\\
\cline{2-5}
& \makecell[c]{\cite{Caching2}} & \makecell[c]{Content} & \makecell[c]{Caching source } &  \makecell[c]{Cooperative hierarchical framework} \\
 \hline
{\makecell[c]{Deployment}} & \makecell[c]{\cite{DeploymentAccess}}  & \makecell[c]{Location + Energy}  & \makecell[c]{Location deployment} & \makecell[c]{Convex optimization}\\
 \hline
\end{tabular}
\end{center}
\label{metirc}
\end{table*}
\subsection{Spectrum Market}
The main purpose of small cell networks is to provide higher channel rate and improve spectrum efficiency through dense deployment. With the increasing traffic demands and users, the spectrum resource is showing shortage. Compared with the traditional static and fixed management and dynamic open cognitive radio, the spectrum market mechanism assigns resource based on users' price. It keeps a dynamic and economical assignment and, avoids the disorder caused by random access. Hence, the spectrum market summarizes advantages from the traditional and cognitive radio management, and achieves  effective results.

The CAGB has been used in some categories of spectrum market, such as auction\cite{Groupon,SunJSAC,YangdejunTVT}. In these cases, users may group to increase the access probability through aggregating their budgets and reduce the cost through making full use of the spectrum. Among these research, the most common awareness is demand awareness. The economical evaluation of spectrum resource is not only related to communication metrics but also to demands and budgets. The demand awareness is also the key point during the shifting from quality of service to quality of experience.

\subsection{Load Balancing}
Since the traffic does not distribute uniformly, the cell resource cannot always match the traffic demand. There always comes the situation where one cell is overloaded but the nearby cells still have extra resource. Load balancing \cite{Loadbalance} is proposed to solve this problem by balancing the traffic and the demand among cells. Some other technologies also employ the same idea with different purposes. For example, sleep mode \cite{Sleepmode} focuses on the energy efficiency and would like to shut down some cells. In this way, the energy is saved by closing a cell and its original users are still served well by nearby cells.

In these technologies, there are two levels for grouping: cell level and user level. Nearby cells can form the coalition, and balance the traffic with less cost. On the other hand, users can play together and select cells to achieve better service. How to balance the load effectively and satisfy quality of service constraints is a key point in modeling. Naturally, almost all the work considers about the location awareness. The Poission point process is used to model the distribution of users and cells. Besides, the energy and demand awareness also influence cells' strategies and users' selections.

\subsection{Caching}
Users may have similar traffic demand in SCNs, especially for popular contents. Compared with downloading the same traffic from macro base station every time when users require, the small cells would cache the traffic ahead and provide a better service when users need it. On one hand, the spectrum resource of the repeatedly requested data is saved for other demand. On the other hand, the cached content is delivered directly to users in a more timely manner.

Considering the distance issue, downloading cost from nearby cells might be less than that from macro base station. Since the caching capacity is limited, small cells can reduce the caching cost and improve the spectrum efficiency through cooperative caching \cite{Caching2,WCL2017}. There are some work also extending the caching from cell level to user level. Users form groups, share the traffic through device-to-device communications. Obviously, the related awareness is content awareness, which is key  to model the caching cost and design grouping rules.

There are some other technologies in ultra-dense SCNs related to the CAGB mechanism, for example, the hyper-graph interference management. Due to the fact that traditional binary model cannot describe the summary effect caused by ultra-dense deployment, the hyper-graph model is used to present a more generalized and accurate interference relationship. The users in groups represent hyper-edges and make decision to avoid interference.

Note that the users, small cells and macro base stations have different demands and requirement. Their focus points are also heterogeneous. The corresponding mechanism and algorithm should balance these conflicting goals. The context awareness discussed before is not only about users but also about cells. For example, energy awareness is an important aspect in offloading and servicing macro users for cell consideration. Content awareness is more common for cells' caching issues, which can help cells reduce much resource and improve efficiency. In deployment problems, location cannot be avoided and ignored. Other context awareness also plays important roles when we consider different aspects of the hyper-dense small cell networks. Besides the context awareness, the relationship between users and different cells are complicated. The simple coalition formation rules can contain much complicated relationship, that is hard to describe or formulate. No matter it is cooperative or competitive, coalitional or selfish, with careful utility design and preference order choice, the relationship can be shifted to the coalition formation process. Some previous studies and research investigated the existence of the stable coalition partition and achieved good properties. For example, the Pareto order \cite{Saad2009} can guarantee the existence of the stable coalition partition in systems with finite states. Based on these studies, the context-aware group buying mechanism can makes the complicated relationship between users and cells  simple, in some degree and aspects, fortunately.


\section{Graphical Coalition Formation Games for Context-aware Group Buying}
To employ CAGB in SCNs, a group is needed first. There are many theoretic models related to the cooperation mechanism. Among these models, the coalition formation games are just studying the problems about which users should be cooperated with and the existence of coalition structure. In this section, graphical coalition formation games are introduced to match the resource allocation in SCNs. Then, some differences between theoretical game models and practical communication systems are discussed.
%
\subsection{Graphical CFGs for CAGB}
The coalition formation games \cite{Saad2009,Sunlingyang} study the existence of stable coalition structure. In most CFGs, how to divide the coalition benefit is determined and just focus on the coalition structure. In other words, players' utilities are only related to the joining in or leaving action and there is no bargaining between players among the coalition. This property ensures the coalition formation is more available for some certain strategy situations, such as dynamic spectrum access in spectrum management and user association in load balancing.

\begin{figure}[!t]
\centering
\includegraphics[width=8cm]{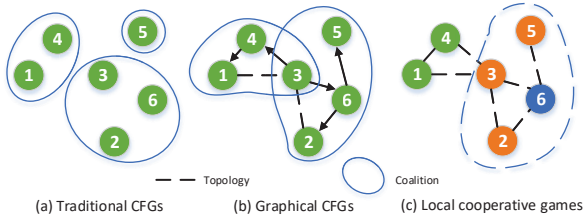}
\caption{Example of tradition CFGs, graphical CFGs and local cooperative games.\small{(Users are all connective to each other in traditional CFGs (a),while the graphical CFGs consider a topology (b). The users in local cooperative games (c) consider all neighbors and its own rewards as utility.)}}
\label{grapggames}
\end{figure}

Compared with the traditional CFGs, some important difference in SCNs are the location and topology. Due to channel fading, users can only connect to nearby cells and users, which limits their coalition selections. Furthermore, utility functions may be related to the network topology, which is also different from the traditional CFGs. The graphical CFGs can model the resource allocation problem in SCNs more accurately while considering the location awareness. An example of the traditional and graphical CFGs is shown in Fig. \ref{grapggames} (a, b).

To match graph in SCNs, measures should be taken from three key points in graphical CFGs: utility function, preference order and coalition actions. The utility function is the objective of players and should be carefully designed first. To evaluate and formulate the influence of topology to users' utilities is necessary and important. Some methods in coalition games and graphical theory may make contribution. Through utilities, preference order can be defined as a selection between two coalitions. According to the preference order, two actions, joining and leaving, determine the merge rule and the split rule of coalition structure. Two coalitions can merge into a larger one when all users achieve benefit from merging. Also, one large coalition can be split into two parts when the divided parts can improve utilities from splitting. The existence of the stable coalition partition is analyzed finally, which is influenced by the utility function and preference order. These actions influence the convergence and the performance of the final results.


There are three key points in the design of CAGB in SCNs: utility function, preference order and coalition action. After introducing the basic definitions, we would discuss how to couple with these challenges.
\begin{itemize}
  \item \textbf{Utility function.} The most challenge in the design of utility function is that how to describe system accurately and provide convergence for proof at the same time. Especially, for the graphical CFGs, the influence of network topology is difficult to formulate in the functions. As a possible way, we can combine the CFGs with other game together, keeping the framework as CFG and utility design as the other games, gathering advantages together to tackle with difficulties. For example, the local cooperative games (LCGs) \cite{XuJSTSP} (in Fig. \ref{grapggames} (c)) as a kind of potential game can offer much help. First, the utility in LCG has a generalized guarantee for NE existence, which provides a large degree of freedom in design. Under this perfect condition, we can consider and formulate the complicated context-aware factors in utility functions, then modify it to couple with LCG utility types for convergence. Secondly, the LCGs can use just local information to achieve the global optimization, which indicates a good adaption for graphical topology and incomplete information situations. With this property, the topology influence can be coupled with and the final performance would have an improvement. Generally, the two properties provide the design convenience, performance guarantee and convergence proof to make the graphical CFGs work. Some other games, such as matching game, Stackelberg game also can be learned from and combined with CFGs.
  \item \textbf{Preference order.} The Pareto order is a common preference order and widely used with a generally good but not optimal performance. Beyond the Pareto order, to propose new orders achieving better performance is another challenge. For example, coalition order maximizing the joint coalition's utility, and selfish order maximizing its own utility without damaging the other mates' utilities in the joint coalition are two new orders. They can describe users' behavior with different objectives or types and achieve better performance than Pareto order. However, without the Pareto order's guarantying, the existence of the final stable coalition partition needs to be proved. The proof is also related to the utility design and some kinds of utilities can provide some insight. For example, the marginal utility, which can be used to describe the coalition utility changes when new users are joining in, has a good adaption for generalized utility formulation and may suit for the coalition order. For the selfish order, some traditional Nash game model can offer some help, due to the similar selfish property in two cases.
  \item \textbf{Coalition action.} Traditional coalition action includes mainly two actions: joining and leaving. Based on these two actions, a swap action is defined as a two-phase action with leaving and joining. Each of the phases should obey the preference order. This requirement keeps the stablity but may not limit the final performance. For example, under the Pareto order, if the leaving would damage other coalition mates' utilities, this behavior would not happen even if the new comer brings a higher benefit. Furthermore, the current swap action considers only one-to-one case, the many-to-many swap mechanism is seldom investigated due to the convergence problem. As the final results under many-to-many action are similar to the Strong Nash equilibrium, it can also provide some help in the proof. To say the least, the new action may also improve the final performance as a heuristic algorithm.
\end{itemize}

\subsection{Gaps Between Game Models and Communication Systems}
While game models provide some methods to allocate resources in SCNs, there are still some gaps between the theoretical game models and the practical communication systems.

The first gap is the information limit. In many game models, the information is assumed to be perfect. The players' strategies, utilities, actions and rewards are easy to get. Then, it is assumed that there are no errors and no delay among these information transmission. However, due to the openness of wireless communication, such information cannot be transmitted without errors. Also, when one player changes its action, the changes cannot be announced to the others immediately. Hence, incomplete and uncertain information is more practical, which brings about difficulties in game model.

Another challenge is the mechanism differences, especially for dynamic factor. Dynamics influences many aspects in communication. Players can move around and their locations are dynamic, which may change the topology of network. For example, the coalition structure cannot be maintained for a long period. The traffic load, channel quality, and other micro dynamics are also varying with time, influencing the utility evaluation. The influence of micro dynamics can be alleviated with utility function design. But for the macro dynamics, some new mechanism or framework need to be proposed to account for the changes.

The final problem is the achieved performance. The game models only provide existence of final equilibriums and the corresponding theoretical analysis. How to approach a better equilibrium of the possible results is still not solved. For the existing learning algorithms, modification is necessary to match system and utility, and the convergence also needs to be proved. Furthermore, considering the above two issues, the learning algorithms are more difficult to design and modify.

\section{Case Study}
In order to motivate the readers, we now present two use cases of context-aware group buying: (1) spectrum auction in \cite{SunJSAC} and (2) cooperative caching, which are related to the situations of Fig. \ref{threeefficiency} (a) and (b), respectively.
\subsection{Case 1: Spectrum Auction}
Spectrum auction is one of the most effective solutions to allocate the spectrum resource, but most of the existing studies assume that the spectrum buyers' demands are homogeneous and the interference relationship is fixed without any change with the variation of spectrum. In this case \cite{SunJSAC}, we combine the overlapping coalition formation with the double auction, jointly considering heterogeneous spectrum reusability, multi-spectrum demand and economical efficiency. We maximize the buyers' benefits and sellers' benefits through the double auction mechanism. Fig. 3 shows the channel selling ratio of different algorithms, which also can be considered as the spectrum efficiency in the system. To search for the optimal coalition structure (i.e., buyer group formation), we present a dynamic and iterative coalition formation algorithm to jointly consider spectrum allocation and pricing rather than separately. Pareto improvement of every coalition operation in the coalitional formation process can make the algorithm finally converge to a stable and satisfactory coalition structure.

\begin{figure}[!t]
\centering
\includegraphics[width=8cm]{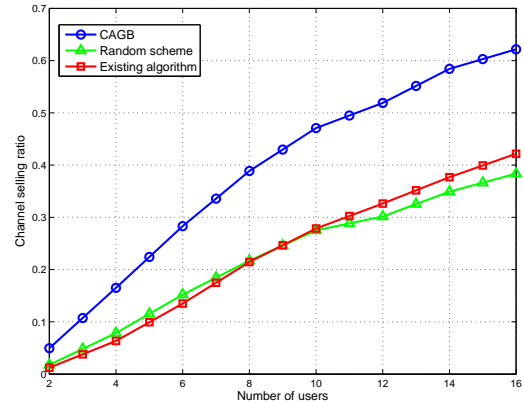}
\caption{Spectrum utilization comparison of CAGB algorithm and others in spectrum auction.}
\label{sunJSAC}
\end{figure}

Fig. \ref{sunJSAC} shows the comparison of spectrum utilization and average buyers' satisfactory levels. The CAGB algorithm is proposed in \cite{SunJSAC} and compared with an existing algorithm and a random solution. The existing algorithm is a truthful double auction for multi-demand heterogeneous spectrum. But it ignores the economic efficiency of the final outcome and just allocates the spectrum for conflict-free buyers in a relatively simple way. Compared with it, we jointly consider the heterogeneous demand, spectrum reusability and economic efficiency in the design of our scheme in \cite{SunJSAC}. We can see that the spectrum utilization and buyers' satisfactory level rise with the increase of the number of buyers. Note that the channel selling ratio (spectrum utilization) is the winning ratio of the whole coalitions. The proposed CAGB algorithm achieves the best performance among the three algorithms.
\subsection{Case 2: Cooperative Caching}
In this case, we study the cooperative caching of small cell networks. We assume that cells have similar contents to cache. Hence, it is better to download the data in groups instead of downloading alone. We focused on users' data cost of downloading from the base station and sharing among the group. We formulated the downloading cost as a Shapley value based on the content awareness of the group data distribution. And the sharing cost is related to the topology within the formed group. The group buying problem with reducing cost is modeled as a coalition formation game. A coalitional order maximizing the coalition benefit, and a selfish order maximizing single user benefit are both proposed. The existence of the stable coalition partitions is also proved in both proposed orders.

\begin{figure}[!t]
\centering
\includegraphics[width=8cm]{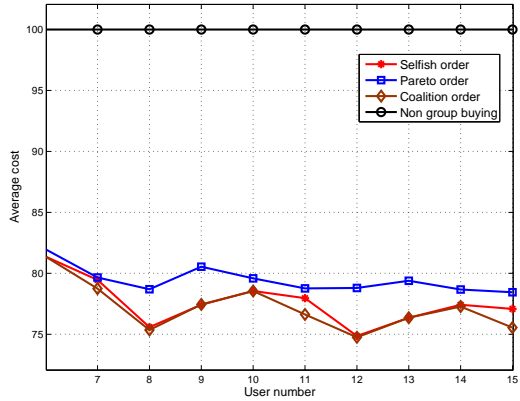}
\caption{The comparison of caching cost with different orders in CAGB and non CAGB situations.}
\label{locationAware}
\end{figure}

The caching cost of users are compared with different orders in Fig. \ref{locationAware}. A non-group buying situation is also simulated for comparison. The three orders, Pareto order, coalition order and selfish order,  are simulated under the CAGB mechanism. From the figure, the three CAGB orders are all better than the non-CAGB situation. The coalition order and selfish order achieve less cost than the Pareto order.
\section{Future Research and Conclusion}
\subsection{Future Research}
Based on the discussion presented in this article, some future work can be focused on the three folds: context awareness, group framework, and buying mechanism.


\begin{itemize}
  \item Enrich and make use of the context awareness widely and deeply. Some other context awareness need to be studied, such as mobility awareness and energy awareness. While the mobility brings about some problems, it also provides another dimension of strategy. Users can move around to join groups, reduce interference and seek for other users' help, to achieve better performance. Furthermore, we can dig the context awareness with a way of big data and deep learning. Abstracting the information to knowledge will contribute to the group buying mechanism to achieve higher performance.
  \item Study a stable and rapid grouping mechanism to match dynamics. The dynamics is one of the most difficult issues in modeling practical systems. The existing works about group formation are most about static situations. However, users have dynamic traffic demand and they are moving. The different dynamics influences the coalition utilities and structure. This requires us to design a rapid and stable coalition formation mechanism to match users' dynamics.
  \item Extend buying mechanism to stereo situations. There are different selling metohds in economics, such as bargaining, pricing and auction. Besides auction, other methods can also be used in group buying mechanisms. The bargaining can model the coalition benefit division among users. Sellers and buyers may use pricing mechanism to compete for more benefit. Besides spectrum resource, energy is yet another potential resource in group buying. A joint spectrum and energy market can be studied in load balancing.
\end{itemize}

\subsection{Conclusion}
In this article, we proposed and discussed a context-aware group buying mechanism for resource allocation in ultra-dense SCNs. Based on the large number of players and proximity location with ultra-dense property, the feasibility, necessity and effectiveness for CAGB mechanism are discussed. A detailed introduction about group buying mechanism is given and some key points of CAGB are presented. Then, some common context awareness are introduced, for example location, demand and content awareness. The relationship between context awareness and group buying are also discussed. Load balancing, spectrum management and cooperative caching are reviewed in view of CAGB. Graphical coalition formation games are presented to model the CAGB. Moreover, one spectrum auction case and one cooperative caching case, both with CAGB, are investigated. Finally, some future research of context awareness, group framework and buying mechanism in CAGB are presented.

\section*{Acknowledgment}
This work was supported by the National Science Foundation of China under Grant No. 61671473, No. 61771488, No. 61631020, the in part by Natural Science Foundation for Distinguished Young Scholars of Jiangsu Province under Grant No. BK20160034, and in part by the Open Research Foundation of Science and Technology on Communication Networks Laboratory.

\ifCLASSOPTIONcaptionsoff
  \newpage
\fi


\begin{thebibliography}{1}
\bibitem{Gexiaohu1}
Xiaohu Ge, Song Tu, Guoqiang Mao, Cheng-Xiang Wang, Tao Han, ``5G Ultra-Dense Cellular Networks,'' \emph{IEEE Wireless Communications}, vol. 23, no. 1, pp.72-79, Feb. 2016.

\bibitem{YangdejunTVT}
Dejun Yang, Guoliang Xue, and Xiang Zhang, ``Group Buying Spectrum Auctions in Cognitive
Radio Networks'', \emph{IEEE Transactions on Vehicular Technology}, vol. 66, no. 1, pp. 810-817, 2017.

\bibitem{SunJSAC}
Youming Sun, Qihui Wu, Jinlong Wang, Yuhua Xu and Alagan Anpalagan, ``VERACITY: Overlapping Coalition Formation-Based Double Auction for Heterogeneous Demand and Spectrum Reusability'', \emph{IEEE Journal on Selected Areas in Communications}, vol. 34, no. 10, pp. 2690-2705, 2016.

\bibitem{Groupon}
Peng Lin, Xiaojun Feng, Qian Zhang and Mounir Hamdi. ``Groupon in the Air: A Three-stage Auction Framework for Spectrum Group-buying,'' \emph{Proceedings IEEE INFOCOM },2013.

\bibitem{Gaolin}
Lin Gao, Lingjie Duan, Jianwei Huang, ``Two-Sided Matching Based Cooperative Spectrum Sharing'', \emph{IEEE Transactions on Mobile Computing}, vol. 16, no. 2, pp. 538-551, 2017.

\bibitem{ZhangruiJSAC}
Jie Xu, Lingjie Duan, and Rui Zhang, ``Energy Group Buying With Loading Sharing for Green Cellular Networks,''
\emph{IEEE Journal on Selected Area in Communications}, vol. 34, no. 4, pp. 786-799, April 2016.

\bibitem{Loadbalance}
Lei You, Di Yuan, ``Load Optimization With User Association in Cooperative and Load-Coupled LTE Networks'', \emph{IEEE Transactions on Wireless Communications}, vol. 16, no. 5, pp. 3218-3231, 2017.

\bibitem{WCL2017}
Yuli Zhang, Yuhua Xu, Qihui Wu, Kailing Yao,  Alagan, Anpalagan, ``A Game-theoretic Approach for Optimal Distributed Cooperative Hybrid Caching in D2D Networks'', \emph{IEEE Wireless Communication Letter, to appear}, 2018.

\bibitem{DeploymentAccess}
Guanhua Qiao, Supeng Leng, Ke Zhang, Kun Yang, ``Joint Deployment and Mobility Management of Energy Harvesting Small Cells in Heterogeneous Networks'', \emph{IEEE ACCESS}, vol. 5, pp. 183-196, 2017.

\bibitem{Gexiaohu2}
Xiaohu Ge, Junliang Ye, Yang Yang, Qiang Li, ``User Mobility Evaluation for 5G Small Cell Networks Based on Individual Mobility Model'', \emph{IEEE Journal on Selected Areas in Communications}, vol. 34, no. 3, pp. 528-541, 2016.

\bibitem{Caching2}
Tuyen X. Tran, Abolfazl Hajisami, Dario Pompili, ``Cooperative Hierarchical Caching in 5G Cloud Radio Access Networks'', \emph{IEEE Network}, vol. 31, no. 4, pp. 35-41, 2017.

\bibitem{Sunlingyang}
Zengfeng Zhang, Lingyang Song, Zhu Han, Walid Saad, ``Coalitional Games with Overlapping Coalitions for Interference Management in Small Cell Networks'', \emph{IEEE Transactions on Wireless Communications}, vol. 13, no. 5, pp. 2659-2669, 2014.


\bibitem{Sleepmode}
Jianchao Zheng, Yueming Cai, Xu Chen, et al., ``Optimal Base Station Sleeping in Green Cellular Networks: A Distributed Cooperative Framework Based on Game Theory.'' \emph{IEEE Transactions on Wireless Communications}, vol. 14, no. 8, pp. 4391-4406, Apr. 2015.

\bibitem{Saad2009}
Walid Saad, Zhu Han, Merouane Debbah, Are Hjorungnes, Tamer Basar, ``Coalitional Game Theory for Communication Networks'', \emph{IEEE Signal Processing Magazine}, Vol. 26, no. 5, pp. 77-97, 2009.

\bibitem{XuJSTSP}
Yuhua Xu, Jinlong Wang, Qihui Wu, Alagan Anpalagan, Yu-Dong Yao, ``Opportunistic Spectrum Access in Cognitive Radio Networks: Global Optimization Using Local Interaction Games'', \emph{IEEE Journal of Selected Topics in Singal Processing}, vol. 6, no. 2, pp. 180-194, 2012.

























\end{thebibliography}
\end{document}